\documentclass[aps,pra,floatfix,twocolumn]{revtex4}
\usepackage{amsmath}
\usepackage{bm}
\usepackage{epsfig}

\def\be{\begin{equation}}
\def\ee{\end{equation}}
\def\bea{\begin{eqnarray}}
\def\eea{\end{eqnarray}}

\begin{document}
\title{{\Large Controlling Irreversibility and Directional Flow of Light with
Atomic Motion }}
\author{C. H. Raymond Ooi}
\affiliation{\sl Department of Physics, Korea University,
Anam-dong, Seongbuk-gu, Seoul, 136-713 Republic of Korea}
\date{\today}

\begin{abstract}
The Doppler effect of moving atoms can create irreversibility of light. We
show that the laser field in electromagnetic induced transparency (EIT)
scheme with atomic motion can control the directional propagation of two
counter-propagating probe fields in atomic gas cell. The quantum coherence
effect serves as an optical transistor. Interference of the two output
fields from the cell shows useful feature for determining the mean atomic
velocity and can be useful as quantum velocimeter. We also find that the
sign of the dispersive phase in EIT has a unique property, which helps to
explain certain features in the interference.
\end{abstract}

\maketitle

In the presence of matter, the reversibility of light can be
broken. An example of the broken symmetry is between the
absorption and emission processes, due to the presence of
spontaneous emission. Atomic motion also breaks the symmetry
between absorption and stimulated emission, which is one of the
ingredients for laser cooling \cite{laser cooling}. No doubt,
irreversibility of light is possible. But how can we control the
irreversibility and incorporate it for useful applications? In a
two-level atom, a negative detuned laser would be resonantly
absorbed if the atom moves opposite to it and would be transmitted
if the atom moves along it, but there is no way to control it.

\begin{figure}[tbp]
\center\epsfxsize=7.5cm\epsffile{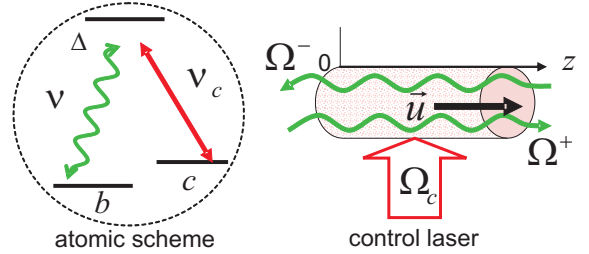}
\caption{(Color online) A gas with atoms with mean velocity $\bar{u}$ in
three-level EIT scheme. The transparency of the two counter-propagating
probe fields is controlled by $\Omega _{c}$.}
\label{scheme}
\end{figure}

In this paper, we consider an ensemble of atoms with three-level $\Lambda $
or electromagnetic induced transparency (EIT) scheme moving with a velocity $%
\vec{u}$,\ sufficiently large such that the Doppler shift is greater than
the linewidth of the excited state, $ku>\Gamma $. In each atom (Fig. \ref
{scheme}), the same transition ($a\leftrightarrow b$) couples to \emph{two
counter-propagating} probe fields with Rabi frequencies $\Omega ^{+}$ and $%
\Omega ^{-}$, which are typically weaker than the control field $\Omega _{c}$%
. The present scheme should be discerned from the velocity selective
coherent population trapping (VSCPT) \cite{VSCPT} which also has
counter-propagating fields, but only one field couples to each transition.

The EIT scheme has been widely studied; mainly from the
perspectives of slow light, nonlinear processes and information
storage \cite{EIT review}. The effect of Doppler broadening on
linewidth \cite{EIT Javan} and slow light \cite{EIT Olga} have
been considered. However, potential application based on
irreversibility of two counter-propagating probe fields in moving
atoms has not been considered. Under certain controllable
condition, one of the probe fields can be made transparent while
the other is strongly absorbed. The combined effect of the center
of mass motion and the control field in EIT provides controllable
$\emph{optical}$ $\emph{rectification}$ to the two
counter-propagating fields. Based on this mechanism, we discuss
possible applications as \emph{optical transistor} and a
\emph{quantum velocimeter}. Further analysis also yield insights
on the difference between the detuning from a real level and from
a level split by the control laser field.

The scheme in Fig. \ref{scheme} is described by the interaction Hamiltonian $%
V=-\hbar \lbrack |a\rangle \langle c|\Omega _{c}e^{-i\nu _{c}t}+|a\rangle
\langle b|(\Omega ^{+}e^{ik\hat{z}}+\Omega ^{-}e^{-ik\hat{z}})e^{-i\nu t}+$%
adj.$]$ with the bare Hamiltonian composed of the kinetic energy $H_{kin}=%
\tfrac{\mathbf{\hat{P}}^{2}}{2M}$ and atomic energy $H_{a}=\sum%
\limits_{l=e,g_{1},g_{2}}\hbar \omega _{l}|l\rangle \langle l|$. The control
laser is taken to be orthogonal to the atomic beam and has no center of mass
(c.m.) effect along the z-axis. Only the counter-propagating probe lasers
give the c.m. effect. The center of mass position operator $\hat{z}$ is
quantized and this gives rise naturally to the first order Doppler shift $%
\omega _{p}=kp/M$ and recoil shift $\omega _{r}=\hbar k^{2}/2M$ in the
transition frequency.

We have a set of infinite number of density matrix equations which contain
finite coherence between different momentum families $\rho _{\alpha \beta
}(p_{\alpha },p_{\beta })=\langle \alpha ,p_{\alpha }|\hat{\rho}|\beta
,p_{\beta }\rangle $ that can only be solved numerically. The presence of
quantized c.m. motion makes it impossible to find exact solutions. However,
we find that it is possible to obtain analytical results in the weak field
limit.

For sufficiently weak probe fields as in typical EIT case, the population in
the excited state is negligible. Thus it is a good approximation to
disregard the momentum redistribution as the result of spontaneous emission,
which gives rise to an integral terms \cite{c.m. spe} in the equations for
the populations. Also, the density matrix equations may solved analytically
by truncating the set of equations based on the approximation of neglecting
the coherences between two momentum families with $2\hbar k$ and larger,
i.e. $\rho _{bc}(p\pm 3\hbar k,p)<<\rho _{bc}(p\pm \hbar k,p),\rho
_{aa}(p\pm 2\hbar k,p)<<\rho _{aa}(p,p)=\rho _{aa}(p),\rho _{bb(cc)}(p-\hbar
k,p+\hbar k)<<\rho _{bb(cc)}(p)$. We then obtain nine equations in quasi
steady state 
\begin{equation}
T_{ab}^{\pm }A^{\pm }(p)\simeq iB^{\pm }(p)\Omega _{c}-i\Omega ^{\mp
}w_{ab}^{\pm }(p)  \label{pab+-}
\end{equation}
\begin{equation}
T_{bc}^{\pm \ast }B^{\pm }(p)\simeq i\Omega _{c}^{\ast }A^{\pm }(p)-i\Omega
^{\mp }C(p)  \label{pcb+-}
\end{equation}
\begin{equation}
T_{ac}^{\ast }C(p)\simeq i\Omega _{c}^{\ast }w_{ac}(p)-i\Omega ^{+\ast
}B^{-}(p)-i\Omega ^{-\ast }B^{+}(pt)  \label{pca}
\end{equation}
where the slowly varying coherences are
\begin{eqnarray}
A^{\pm }(p,t) &=&e^{i(\Delta \pm \omega _{p}+\omega _{r})t}\rho _{ab}(p,p\pm
\hbar k,t),  \label{A} \\
B^{\pm }(p,t) &=&e^{i(\Delta -\Delta _{c}\pm \omega _{p}+\omega _{r})t}\rho
_{cb}(p,p\pm \hbar k,t)  \label{B} \\
C(p,t) &=&e^{-i\Delta _{c}t}\rho _{ca}(p,t)  \label{C}
\end{eqnarray}

The complex decoherences that include the Doppler and recoil shifts are $%
T_{ab}^{\pm }=\gamma _{ab}-i(\Delta \pm \omega _{p}+\omega _{r})$, $%
T_{bc}^{\pm }=\gamma _{bc}-i(\Delta _{c}-(\Delta \pm \omega _{p}+\omega
_{r}))$ and $T_{ac}=\gamma _{ac}-i\Delta _{c}$. The population inversions
are defined as $w_{ac}(p)=\rho _{aa}(p)-\rho _{cc}(p)$ and $w_{ab}^{\pm
}(p)=\rho _{aa}(p)-\rho _{bb}(p\pm \hbar k)$ which can be taken to be the
initial values for weak probe fields $\rho _{aa}(p,t)\simeq \rho
_{aa}(p,0)=\rho _{aa0}f(p)$, $\rho _{cc}(p,t)\simeq \rho _{bb}(p,0)$ and $%
\rho _{bb}(p\pm \hbar k,t)\simeq \rho _{bb}(p,t)\simeq \rho _{bb}(p,0)=\rho
_{bb0}f(p)$. The populations depend on the momentum distribution of a gas
jet $f(p)=\exp [-(p-\bar{p})^{2}/\Delta p^{2}]$ with $\int_{-\infty
}^{\infty }f(p)dp=1$. The gas has velocity width of $\Delta p=\sqrt{2Mk_{B}T}
$ and a mean momentum $\bar{p}=m\bar{u}$.

The five coupled (\ref{pab+-})-(\ref{pca}) can be solved exactly
\begin{eqnarray}
A^{\mp }(p,z) &=&-i\frac{\Omega ^{\pm }}{\Upsilon }[I^{\mp }(w_{ab}^{\mp
}T_{ab}^{\pm }T_{bc}^{\ast \mp }+w_{ab}^{\pm }I_{c})/J^{+}J^{-}+  \notag \\
&&w_{ab}^{\mp }I^{\pm }/J^{\mp }+(w_{ab}^{\mp }T_{ac}^{\ast }T_{bc}^{\ast
\mp }-w_{ac}I_{c})/J^{\mp }]  \label{A nonlinear}
\end{eqnarray}
where $\Upsilon =T_{ac}^{\ast }+I^{-}T_{ab}^{p}/J^{p}+I^{p}T_{ab}^{-}/J^{-}$%
, $J^{\pm }=T_{ab}^{\pm }T_{bc}^{\ast \pm }+I_{c}$ and $I^{\pm }=|\Omega
^{\pm }|^{2}$, $I_{c}=|\Omega _{c}|^{2}$.

The solutions are related to the propagation equations
\begin{equation}
\frac{\partial }{\partial z}\Omega ^{\pm }(z)=i\kappa g\int_{-\infty
}^{\infty }A^{\mp }(p,z,I^{+},I^{-})dp=G^{\pm }\Omega ^{\pm }  \label{propg}
\end{equation}
\ where $\kappa g=N\frac{\mu _{o}c\omega |\wp |^{2}}{2\hbar }=N\frac{\omega
|\wp |^{2}}{2\hbar \varepsilon _{0}c}$. Please note that we have done the
momentum integration differently from previous works \cite{EIT Olga}, \cite
{EIT Javan}. Note that this is a general approach, that enables the
populations in different levels to take different distributions; for
example: $\rho _{cc}(p)=0.2f(p),\rho _{bb}(p)=0.8f(p-\hbar k),\rho
_{aa}(p)=0 $ subjected to normalization $\sum\limits_{x=a,b,c}\int_{-\infty
}^{\infty }\rho _{xx}(p)dp=1$.

Equations (\ref{A nonlinear})-(\ref{propg}) which imply that the variables $%
A^{\pm }$, $B^{\pm }$ and $C$ depend on $z.$ The equations are highly
nonlinear due to the dependence of Eq. (\ref{A nonlinear}) on $I^{\pm }$,
but can be solved numerically. The nonlinearity and cross-coupling of the
fields arise from the last two terms in Eq. (\ref{pca}). The cross-coupling
correspond to two probe fields interacting simultaneously with the same atom.

In the limit of small probe fields $I^{\pm }<<I_{c},(T_{ab}^{\pm })^{2}$ the
cross-interaction is weak and negligible, and we have a linear theory $%
A^{\pm }(p,z)=\frac{-i\Omega ^{\mp }}{J^{\pm }}[w_{ab}^{\pm }T_{bc}^{\ast
\pm }-w_{ac}I_{c}/T_{ac}^{\ast }]$ and Eqs. (\ref{propg}) yield the
solutions $\Omega ^{\pm }(z)=\Omega ^{\pm }(0)e^{G^{\pm }z}$ where the
complex 'gain' becomes
\begin{equation}
G^{\pm }=\kappa g\int \frac{w_{ab}^{\mp }T_{bc}^{\ast \mp
}-w_{ac}I_{c}/T_{ac}^{\ast }}{J^{\mp }}dp  \label{G+-}
\end{equation}
If we neglect the superscripts $\pm $, we recover the known \cite{EIT Javan}
relation $\tilde{\rho}_{ab}\simeq -i\Omega \frac{w_{ab}T_{bc}^{\ast
}-w_{ac}I_{c}/T_{ac}^{\ast }}{(T_{bc}^{\ast }T_{ab}+I_{c}\allowbreak )}$.

\begin{figure}[tbp]
\center\epsfxsize=8.5cm\epsffile{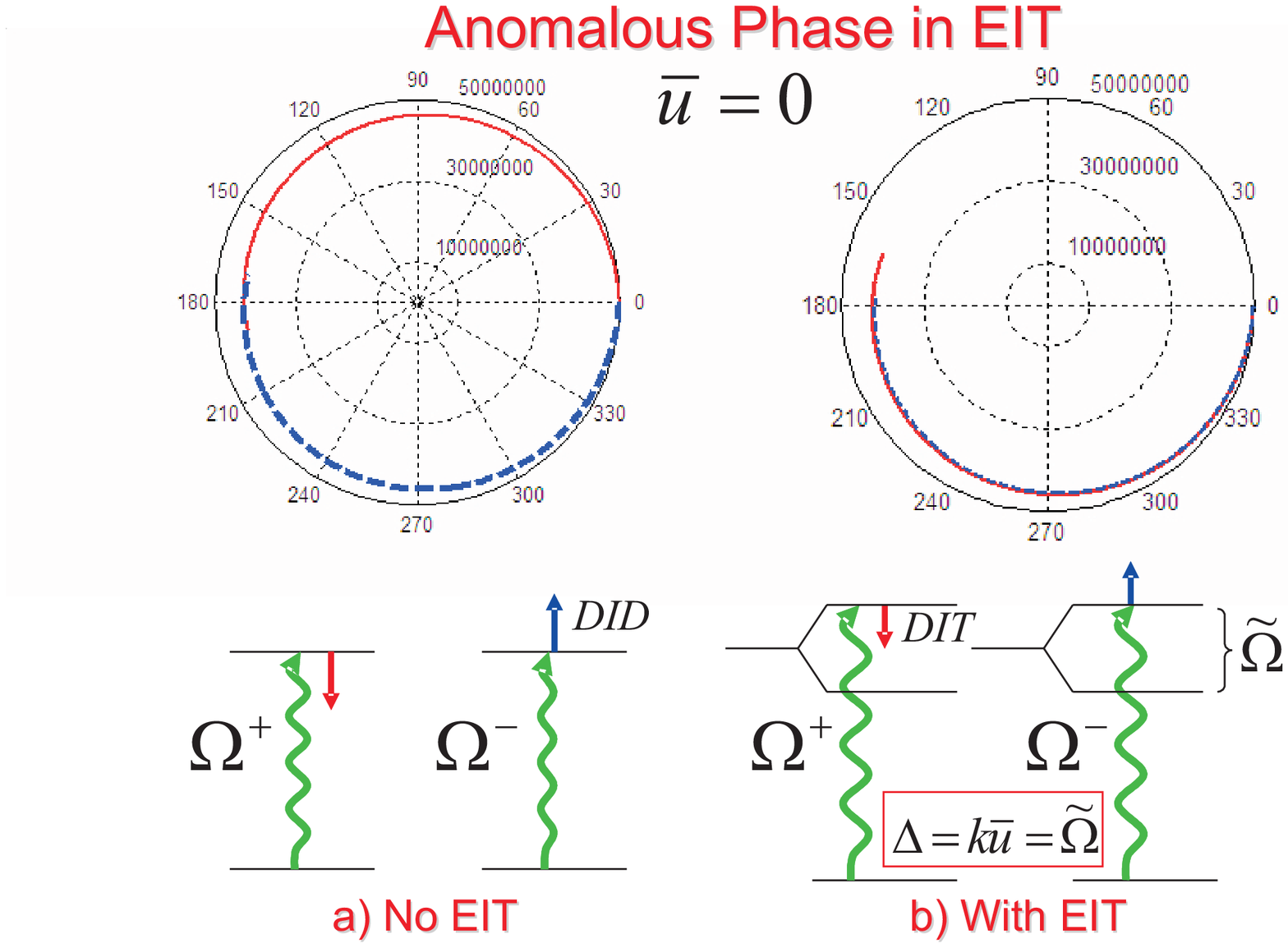}
\caption{(Color online) Spatial variations of the probe fields
$|\Omega ^{+}| $(red solid line) and $|\Omega ^{-}|$(blue dash
line) through polar
plots for : a) $\Delta =0,\Omega _{c}=0$, b) $\Delta =k\bar{u}$ and finite $%
\Omega _{c}$ such that $\tilde{\Omega}(=\protect\sqrt{\Omega _{c}^{2}-%
\protect\gamma _{ac}^{2}})=\protect\omega _{D}$. The upper level $a$ is
split by $2\tilde{\Omega}$. Doppler shifts of co-propagating field (red
arrow) and counter-propagating field(blue arrow) lead to Doppler induced
detuning (DID), Doppler induced transparency(DIT) and possibly Doppler
induced resonant (DIR). We use mean velocity $\bar{u}=300$ms$^{-1}$,
Gaussian width $\Delta u=\protect\sqrt{k_{B}T/M}$, $T=1K$, $M=87$(Rb), $%
\protect\gamma _{bc}=0.1\protect\gamma _{ac}$ and propagation length $L=0.1m$%
.}
\label{Omvszpolarboth}
\end{figure}

The real parts of $G^{\pm }$ give the absorption coefficients (if negative)
and gain (if positive) while the imaginary part gives the the change in the
wavevector. We plot $|\Omega ^{\pm }(z)|=\Omega _{0}e^{\text{Re}G^{\pm }z}$
in Fig. \ref{Omvszpolarboth} with the assumption of two-photon resonance $%
\Delta =\Delta _{c}$ and decoherence $\gamma _{bc}=0.1\gamma _{ac}$. The
polar plots illustrate the spatial evolutions of the amplitudes $|\Omega
^{\pm }(z)|$ (radial lengths) and the phases, $\theta ^{\pm }=$Im$G^{\pm }z$
(angles). In the absence of atomic motion, we find a new subtle effect of
EIT. The polar plot (Fig. \ref{Omvszpolarboth}b) shows that linear responses
or the susceptibilities $\chi ^{(1)\pm }$ for the two fields $\Omega ^{\pm }$
are the same when the control field is on, which is in contrast to the case
without control field (Fig. \ref{Omvszpolarboth}a) where the phases of the
two fields have opposite signs.

\emph{Directional Propagation}

\begin{figure}[tbp]
\center\epsfxsize=8.5cm\epsffile{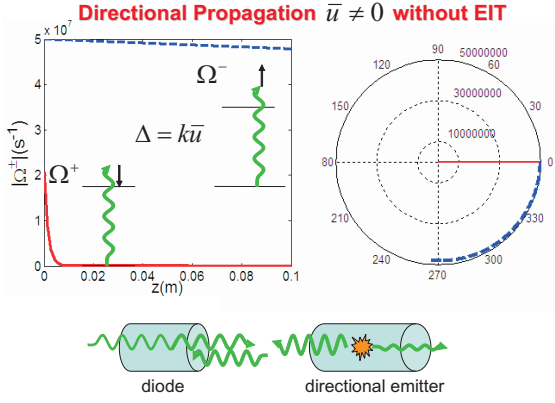}
\caption{(Color online) For positive detuning $\Delta >0$ the
field $\Omega ^{+}$ propagating to the right is red-shifted by
atoms to resonant and heavily absorbed. The mean velocity is
towards the right $\bar{u}>0$. No control field is available. If
the probe is passed through, the medium works like an optical
diode. If the signal is generated from within the medium, it works
like a directional emitter.} \label{directionality}
\end{figure}

We start by analyzing the simplest case of without EIT ($\Omega _{c}=0$)
shown in Fig. \ref{directionality}. When the probe is resonant $\Delta =0$,
both counter-propagating fields are equally detuned (with opposite signs)
from the upper level by the Doppler shift (neglect recoil shift). If the
probe is detuned $\Delta =-\omega _{\bar{p}}$ ($<0$), the field that
propagates opposite to the atoms would be absorbed since $G^{-}\simeq \kappa
gw_{ab0}\mathcal{GL}$ is real and negative-value where $\mathcal{GL}%
=\int_{-\infty }^{\infty }\frac{e^{-x}}{\gamma ^{2}+x^{2}}dx,x=\omega
_{p}-\omega _{\bar{p}}$ while the co-propagating field is transmitted since $%
G^{+}\simeq -i\kappa gw_{ab0}\int \frac{f(p)dp}{\omega _{\bar{p}}+\omega _{p}%
}$ is primarily imaginary for $\gamma <<\omega _{\bar{p}}$. The
counter-propagating probe fields undergo \emph{optical rectification}, i.e.
one of the probes is transparent while the other probe field is absorbed.
Thus, an appropriate probe field detuning with respect to the mean velocity
of a gas jet can serve as an \emph{optical diode}.

The mechanism may also be used as a \emph{directional emitter} if an
emitting source like an electrically driven quantum dot is embedded inside
the channel of hollowed waveguide or fiber containing a gas flow. The source
would emit light and would be guided along two opposite directions. The
light in one direction would propagate with little damping while the light
in the opposite direction is heavily absorbed.

\emph{Control of Directionality}

\begin{figure}[h]
\center\epsfxsize=8.5cm\epsffile{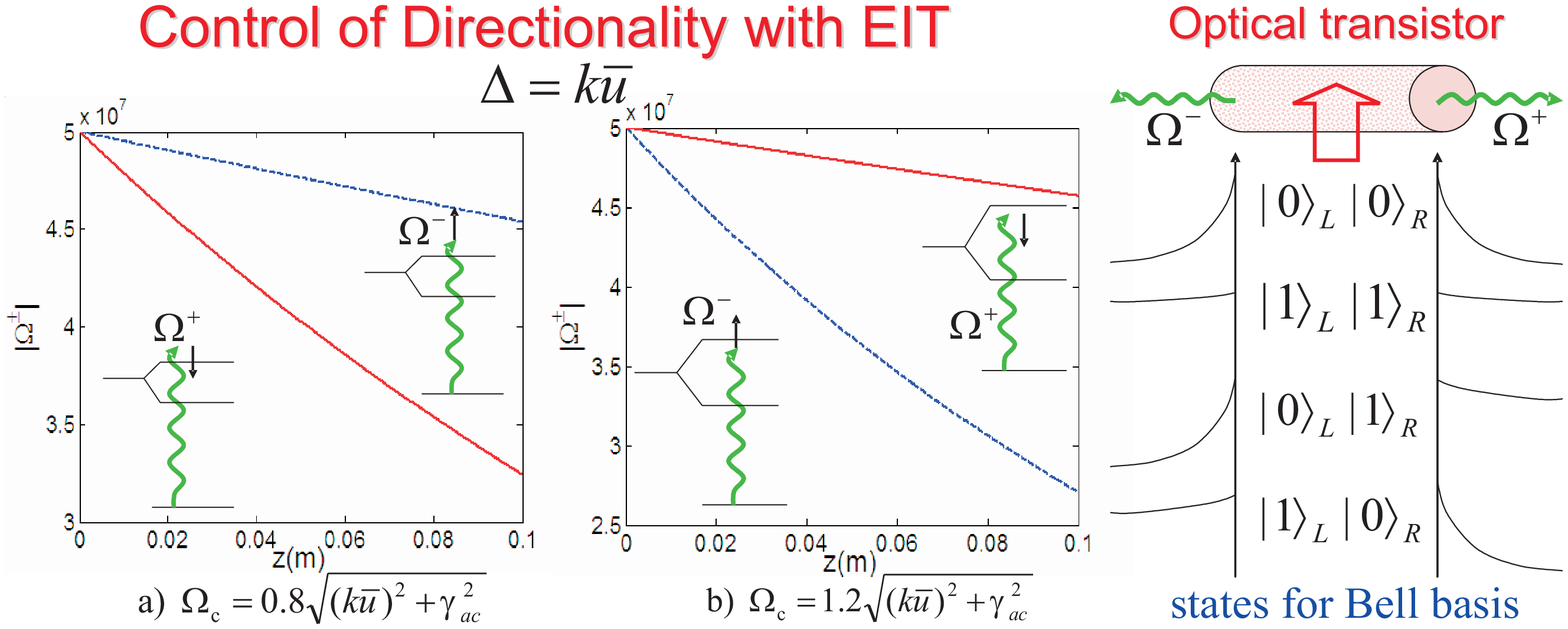}
\caption{(Color online) a) and b) show how one of the fields can
be damped greater than the other by controlling the laser field.
c) Optical transistor with two outputs produces four distinct
states based on transparency and damping. }
\label{controltransistor}
\end{figure}

Now, by applying a laser field $\Omega _{c}$ in EIT configuration, we can
control the directional emission by controlling the absorption strength of
the two fields. The field $\Omega _{c}$ splits the upper levels by $2\tilde{%
\Omega}=2\sqrt{\Omega _{c}^{2}-\gamma _{ac}^{2}}$, while the detuning which
may add up or subtract with the Doppler shift. For $\tilde{\Omega}<\omega _{%
\bar{p}}$, the co-propagating field is damped more rapidly than the
counter-propagating field because of its proximity to one of the split
levels upon the Doppler shift. The reverse applies for $\tilde{\Omega}%
>\omega _{\bar{p}}$(Figs. \ref{controltransistor}a and b). In the case of $%
\tilde{\Omega}=\omega _{\bar{p}}$ , the two fields are equally detuned from
the ac Stark shifted upper level.

\emph{Optical Transistor }

In principle, the EIT scheme with stationary atoms works like an \emph{%
optical valve}, with the control field acting as a knob that controls the
intensity of the transmitted probe signal. In the absence of \emph{atomic
motion} or Doppler effect, both counter-propagating fields would be either
equally damped or transparent but there is no \emph{rectification}. The
presence of atomic motion creates rectification of the probe signals, by
damping one of the signals and inducing a transparency on the other signal.
The ability of differential control of the two counter-propagating fields
makes the device like an \emph{optical transistor} as shown in Fig. \ref
{controltransistor}c. The two signals to be rectified, transmitted or
blocked through the flow of fluid and the control field.

\emph{Quantum Velocimeter}

\begin{figure}[h]
\center\epsfxsize=8.5cm\epsffile{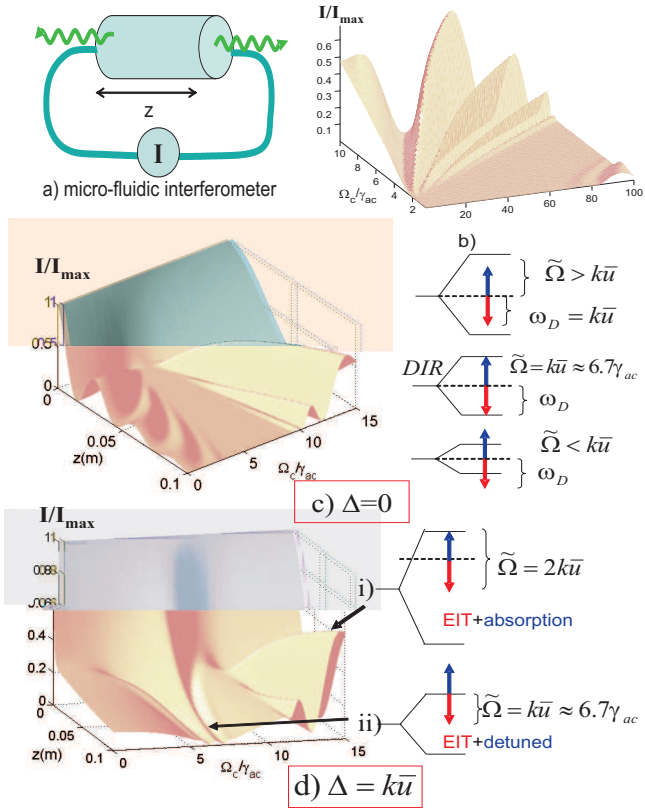}
\caption{(Color online) a) opto-fluidic interferometer, b) sensitivity of
the device is found from the interference signal (normalized) versus the
control field $\Omega _{c}$ and mean velocity $\bar{u}$ at $z=0.1m$. The
mean velocity is obtained from the variation of the interference with $z$
and $\Omega _{c}$ for c) $\Delta =0$ and d) $\Delta =k\bar{u}$, where $\bar{u%
}=100ms^{-1}$ corresponds to $\protect\omega _{D}/\protect\gamma _{ac}=6.67$%
. }
\label{interf}
\end{figure}

An interferometer setup with one arm in a quantum coherence medium has been
proposed as magnetometer \cite{magnetometer}. We propose a new different
setup (Fig. \ref{interf}a) where \emph{both} channels are in the phaseonium
(EIT) medium and sensitive to Doppler effect. The interference of the
counter-propagating fields can be used to detect atomic motion in a gas. The
use of Doppler effect to measure the atomic velocity reminds us of the
existing technique of Laser Doppler Velocimetry \cite{velocimetry} which
uses cross laser beam and Doppler effect to measure the flow velocity, a
concept entirely based on classical physics. Here, we introduce \emph{%
quantum velocimetry} that incorporates a different underlying mechanism,
based on quantum coherence effect through the EIT with a laser as a control
knob to create a sensitive velocimeter.

By combining the two probe fields as shown in Fig. \ref{interf}a) the total
field becomes
\begin{equation}
\Omega ^{tot}(z)=\Omega ^{+}(z)+\Omega ^{-}(z)=\Omega
_{0}(e^{G^{+}z}+e^{G^{-}z})
\end{equation}
The presence of atomic motion in the gas gives rise to $G^{+}$ $\neq G^{-}$.
The real part gives the absorption or transparency of the field. The
imaginary part corresponds to the linear susceptibility $\chi ^{\pm }=gN\wp
\tilde{\rho}_{ab}(p,p\mp \hbar k,t)/\varepsilon _{0}\Omega ^{\pm }$ and it
gives oscillations or beating in $I(z)=|\Omega ^{tot}(z)|^{2}$, as shown in
Fig. \ref{interf}a. The key point is that atomic motion can be detected
through the presence of oscillations in $I(z)$ versus $z$, since there are
no oscillations when $\bar{u}=0$.

The sensitivity corresponds to the ability to detect a small change in
velocity $\delta u=\frac{\Delta u}{\Delta \Omega _{c}}\delta \Omega _{c}$.
From Fig. \ref{interf}b, we estimate that $\frac{\Delta u}{\Delta \Omega _{c}%
}\sim 15/\gamma _{ac}$ which corresponds to $\Delta \omega _{D}=k\Delta \bar{%
u}\sim \Delta \Omega _{c}$. Thus, the classical sensitivity is
\begin{equation}
\delta u=\lambda \delta \Omega _{c}/2\pi  \label{du}
\end{equation}
which depends on the probe wavelength and the smallest detectable variation
in the intensity. By taking $\delta \Omega _{c}\sim 3\times 10^{4}$s$^{-1}$
as the smallest detectable value and $\lambda \sim 2\times 10^{-6}m$, we
find $\delta u=10mm/s$, which is enough to measure a flow as slow as the
speed of an ant.

For $\Delta =0$ Fig. \ref{interf}c shows a valley or channel feature which
decays exponentially at around $\Omega _{c}\simeq \omega _{D}=k\bar{u}$.
Here, the two fields are shifted to resonance and the imaginary part of $%
G^{\pm }$ vanishes, so no oscillations. The probe fields add in the form $%
(e^{-bz}+e^{-bz})^{2}\symbol{126}e^{-2bz}$. The channel divides the profile
into two regions with oscillations. The oscillations in $\Omega _{c}<k\bar{u}
$ are more rapid than the region $\Omega _{c}>k\bar{u}$. The fields in these
regions interfere as $(e^{iaz}e^{-bz}+e^{-iaz}e^{-bz})^{2}\sim e^{-2bz}\cos
^{2}az$.

For finite detuning (Fig. \ref{interf}d), the ridge at $\Omega _{c}\simeq
\omega _{D}$ corresponds to equal detuning (but opposite signs) of both
probe fields from the ac Stark shifted level. One experiences EIT while the
other is just normal detuning. We learn from Fig. \ref{Omvszpolarboth}b that
their phases would be the same sign a superposition of EIT. So the fields
interfere as $|e^{-bz}e^{iaz}+e^{-bz}e^{-iaz}|^{2}\rightarrow e^{-2bz}$,
which explains the exponential decay. However, at $\Omega _{c}\simeq 2\omega
_{D}$, we have one EIT and one resonant with the shifted upper level. Here,
the fields interfere as

\begin{equation}
|e^{-bz}e^{iaz}+e^{-(b+c)z}|^{2}=e^{-2bz}[4e^{-cz}\cos ^{2}\frac{az}{2}%
+(e^{-cz}-1)^{2}]  \label{analytic}
\end{equation}
where $c>0$ which means that the damping due to absorption is larger than
the damping in EIT. The rate of oscillations is less rapid by half. The
rising ridge can be explained as due to the term $(e^{-cz}-1)^{2}$ in Eq. (%
\ref{analytic}).

Before we conclude, we briefly discuss a possible connection or analogy
between the classical directional states with quantum states. For $\bar{u}=0$%
, a resonant field with $\Omega _{c}=0$ is damped, giving no-field state $%
|0\rangle _{L}|0\rangle _{R}$. However, a resonant field with finite $\Omega
_{c}$ corresponds to transparency (EIT), thus the state of light is $%
|1\rangle _{L}|1\rangle _{R}$. For positive $\bar{u}$, when $\Delta =k\bar{u}
$ co-propagating field experiences transparency whereas the
counter-propagating field is absorbed, thus the state is $|0\rangle
_{L}|1\rangle _{R}$, refered as ''right-field''. For $\Delta =-k\bar{u}$ and
$\tilde{\Omega}=2k\bar{u}$ or (for negative $\bar{u}$ and $\Delta =k|\bar{u}|
$) the situation is reversed giving the ''left-field'' state $|1\rangle
_{L}|0\rangle _{R}$. The states are subjected to physical factors, namely
the driving field and the direction of the fluid. If these four states
(shown in Fig. \ref{controltransistor}c) can be used to construct the
well-known Bell basis the scheme could be useful in quantum information. For
example, the superposition of on and off control field can be described by a
macroscopic entangled state
\begin{equation}
|\Phi ^{\pm }\rangle =|0\rangle _{L}|1\rangle _{R}\pm |1\rangle
_{L}|0\rangle _{R}
\end{equation}
Similarly the superposition of ``right-field'' and ``left-field'' states is
described by
\begin{equation}
|\Psi ^{\pm }\rangle =|0\rangle _{L}|1\rangle _{R}\pm |1\rangle
_{L}|0\rangle _{R}
\end{equation}
which represents an indefinite mean velocity associated with the case of
chaotic flow.

Finally, we conclude by noting that the full potential of optical
directional control could be realized through miniaturization. The proposed
optical transistor and velocimeter can be integrated with the existing
microfluidic technology \cite{microfluidic}, creating a new class of \emph{%
optical-microfluidic} sensor for further advances in chemical and biological
sensing.



\begin{thebibliography}{99}
\bibitem{laser cooling}  C. Cohen-Tannoudji, Rev. Mod. Phys. \textbf{70},
707 (1998); W. D. Phillips, \textit{ibid.} \textbf{7}0, 721 (1998); S. Chu,
\textit{ibid.} \textbf{70}, 685 (1998); J. Dalibard and C. Cohen-Tannoudji,
J. Opt. Soc. Am\textit{.} \textbf{6}, 2023 (1989); T. Hansch and A.
Schawlow, Opt. Comm. \textbf{13}, 68 (1975);V. Letokhov, JETP Lett. \textbf{7%
}, 272 (1968).

\bibitem{VSCPT}  O. N. Prudnikov and E. Arimondo, J. Opt. Soc. Am. B \textbf{%
20}, 909 (2003); A. Aspect et al, Phys. Rev. Lett. \textbf{61}, 826 (1988).

\bibitem{EIT review}  M. Fleischhauer, A. Imammoglu, J. P. Dowling, Rev.
Mod. Phys. 77, 633 (2005).

\bibitem{EIT Javan}  A. Javan, O. Kocharovskaya, H. Lee, and M. O. Scully,
Phys. Rev. A \textbf{66}, 013805 (2002).

\bibitem{EIT Olga}  O. Kocharovskaya, Y. Rostovtsev, and M. O. Scully, Phys.
Rev. Lett. \textbf{86}, 628 (2001).

\bibitem{c.m. spe}  C. H. Raymond Ooi, K.-P. Marzlin and J. Audretsch, Phys.
Rev. A \textbf{66}, 063413 (2002).


\bibitem{magnetometer}  M. O. Scully and M. Fleischhauer, Phys. Rev. Lett.
\textbf{69}, 1360 (1992); M. Fleischhauer and M. O. Scully, Phys. Rev. A
\textbf{49}, 1973 (1994).

\bibitem{velocimetry}  T. D. Fansler and D. T. French, Appl. Opt. \textbf{32}%
, 3846 (1993); L. E. Drain, \textit{The Laser Doppler Technique}, Wiley,
Chichester (1980).

\bibitem{microfluidic}  T. M. Squires and S. R. Quake, Rev. Mod. Phys.
\textbf{77}, 977 (2005).
\end{thebibliography}
\end{document}